\documentclass[a4paper,12pt]{article}
\usepackage{amsmath,amsfonts,amssymb}

\begin{document}

\begin{center}
{\Large\bf Large degeneracy in light mesons from a modified soft-wall holographic model}
\end{center}

\begin{center}
{\large S. S. Afonin}
\end{center}

\begin{center}
{\it Saint Petersburg State University, 7/9 Universitetskaya nab.,
St.Petersburg, 199034, Russia}
\end{center}

\begin{abstract}

We propose a modification of soft-wall holographic model that in
combination with the light-front holographic QCD leads to a broad
hydrogen-like degeneracy of the same sort as the degeneracy
observed experimentally in the spectrum of light mesons.
\end{abstract}

\section{Introduction}

The description of spectra of light hadrons is a traditionally difficult
and controversial aspect in the physics of strong interactions.
About fifteen years ago, the gauge/gravity duality in the
string theory inspired the appearance of unusual approach
to hadron physics called holographic QCD or AdS/QCD. In the last decade,
various applications of the holographic approach to the hadron spectroscopy
have been developed in numerous papers. A inherent property of the gauge/gravity duality is
the large-$N$ limit of $SU(N)$ gauge theory.
On the other hand, meson and glueball states become well
defined freely propagating objects in the large-$N$ QCD as their decays and
interactions are suppressed and their masses scale as $\mathcal{O}(1)$ with $N$~\cite{hoof}.
Since the strong coupling phenomena are crucial in the physics of light hadrons,
the idea to apply the holographic methods to description of masses of
light mesons and glueballs looks natural.

The hadron masses appears as poles in correlation functions of currents interpolating
the corresponding hadron states. The holographic method provides a definite prescription
how these correlation functions can be extracted from a dual theory~\cite{witten}. 
On the other hand, the mass spectrum can be found from solution of corresponding equation of motion.
The purpose of the given work is to construct a modified Soft-Wall (SW) holographic model~\cite{son2}
that leads to a broad degeneracy observed experimentally in the spectrum of light mesons.

\section{Spectrum of SW model}

Consider the standard action of the SW model for arbitrary integer spin~\cite{son2},
\begin{equation}
\label{1}
I=(-1)^{J}\frac12\int d^4x\,dz \sqrt{g}\,e^{-az^2}\left(\nabla_N\phi_J\nabla^N\phi^J-
m_J^2\phi_J\phi^J\right).
\end{equation}
The action is written in the AdS$_5$ space. The dilaton background $e^{-az^2}$
allows to get the linear Regge spectrum in which the slope is controlled
by the mass parameter $a>0$. A convenient parametrization
of AdS$_5$ metric is $ds^2=(R^2/z^2)(dx_{\mu}dx^{\mu}-dz^2)$
(the sign is mostly negative and $R$ denotes the AdS radius),
where the holographic coordinate $z$ is interpreted as inverse energy scale.
The point $z=0$ represents thus the UV boundary of the AdS$_5$ space.
The higher spin fields $\phi_J\doteq\phi_{M_1M_2\dots M_J}$, $M_i=0,1,2,3,4$,
are described by symmetric, traceless tensors of rank $J$.
By assumption, the physical components
of 5D fields satisfy the condition,
\begin{equation}
\label{2}
\phi_{z\dots}=0.
\end{equation}
This condition allows to cancel additional
quadratic terms in~\eqref{1} appearing from alternative ways of
contraction of coordinate indices~\cite{son2,katz,br2,br3}.
The constraint $\partial^{\mu}\phi_{\mu\dots}=0$ is also assumed to have
the required $2J+1$ physical degrees of freedom.
The action~\eqref{1} with condition~\eqref{2} describes the $J=0$ and $J=1$ mesons as well.

In the original SW model~\cite{son2}, only the gauge higher-spin fields were considered.
We will discuss a more general case. Since the 5D dual gravitational theory must be
weakly coupled, we may use the "weak gravity" approximation, where the affine
connections in the covariant derivatives $\nabla_N$ are neglected. In this
approximation, $\nabla_N$ are replaced by the usual derivatives $\partial_N$.
In addition, the dependence of 5D mass $m_J$ on spin is completely determined by
the behavior of fields near the UV boundary~\cite{br2,br3},
\begin{equation}
\label{3}
m_J^2R^2=(\Delta-J)(\Delta+J-4),
\end{equation}
where $\Delta$ means the scaling dimension of 4D operator
dual to $\phi^J$ on the UV boundary. Generally speaking,
the mass term entering~\eqref{1} should also depend on the mass parameter $a$
and the holographic coordinate $z$ because of the dilaton background in the action~\eqref{1}.
We will argue that a correct phenomenological description of observed meson spectrum
can be achieved if the 5D mass term is replaced in~\eqref{1} by
\begin{equation}
\label{3b}
m_J^2\longrightarrow m_J^2+2a\left(J-2+\sqrt{(J-2)^2+m_J^2R^2}\right)\left(\frac{z}{R}\right)^2,
\end{equation}
where $m_J^2$ is subject to~\eqref{3}.

The behavior of fields $\phi^J$ near the UV boundary in~\eqref{1}
can be found by extracting the leading $z=0$ asymptotics of "would be"
$z$-dependent vacuum expectation value $\phi_0^J(z)$.
The equation for $\phi_0^J(z)$ follows from the action~\eqref{1},
\begin{equation}
\label{4}
\partial_z\left(e^{-az^2}z^{2J-3}\partial_z\phi_0^J\right)=
(m_J^2R^2+\dots)e^{-az^2}z^{2J-5}\phi_0^J,
\end{equation}
where dots denote the additional $z$-dependent contribution in~\eqref{3b}.
The leading asymptotics at $z=0$ is
\begin{equation}
\label{5}
\phi_0^J\sim z^{2-J-\xi},
\end{equation}
where
\begin{equation}
\label{6}
\xi=\sqrt{(J-2)^2+m_J^2R^2}.
\end{equation}

The asymptotics~\eqref{5} shows that in general case $\phi_0^J(z)$ has no
finite and non-zero value on the UV boundary $z=0$, i.e. $\phi_J(0)$ cannot be interpreted
as a source for some 4D operator in the spirit of AdS/CFT correspondence.
But such an interpretation can be given for a rescaled field $\tilde{\phi}_J$,
\begin{equation}
\label{7}
\phi_J=z^{2-J-\xi}\tilde{\phi}_J.
\end{equation}
We may say that the physical field is the rescaled one $\tilde{\phi}_J$.

Now we are to substitute~\eqref{7} into the action~\eqref{1}, lower indices
and write the equation of motion for $\tilde{\phi}_J$. Making use of the
statndard plane-wave ansatz for physical
particles with 4D momentum $q_n$,
\begin{equation}
\label{8}
\phi_J(x,z)=e^{iq_nx}\phi_n(z)\varepsilon_J,\qquad q_n^2=m_n^2,
\end{equation}
($\varepsilon_J$ is the polarization tensor) we obtain after simple calculations,
\begin{equation}
\label{9}
-\partial_z\left(e^{-az^2}z^{1-2\xi}\partial_z\tilde{\phi}_{n}\right)=
m_{n,J}^2e^{-az^2}z^{1-2\xi}\tilde{\phi}_{n}.
\end{equation}
It should be noted that the leading ultraviolet contributions from the mass terms
(i.e. $\mathcal{O}(z^{-1-2\xi})$ terms in~\eqref{9}) cancel
and the dependence on $m_J$ sits in $\xi$ only. The substitution
\begin{equation}
\label{10}
\tilde{\phi}_{n}=e^{az^2/2}z^{\xi-1/2}\psi_n,
\end{equation}
converts~\eqref{9} into a one-dimensional stationary Schr\"{o}dinger equation,
\begin{equation}
\label{11}
-\partial_z^2\psi_n+V(z)\psi_n=m_{n,J}^2\psi_n,
\end{equation}
\begin{equation}
\label{12}
V(z)=a^2z^2+2a(\xi-1)+\frac{\xi^2-1/4}{z^2}.
\end{equation}
At $\xi>1/2$ it has the discrete spectrum\footnote{If
the sign of $a$ is unspecified, the spectrum is
$m_{n,J}^2=2|a|\left[2n+\left(1+\frac{a}{|a|}\right)\xi+1-\frac{a}{|a|}\right]$.
As was noticed in Ref.~\cite{son3} the spectrum does not depend on spin if $a<0$.}
($n=0,1,2,\dots$)
\begin{equation}
\label{13}
m_{n,J}^2=4a\left(n+\xi\right).
\end{equation}
Substituting~\eqref{3} to~\eqref{6} we have $\xi=|\Delta-2|$.
This yields our final result for the spectrum of SW model,
\begin{equation}
\label{14}
m_{n,J}^2=4a\left(n+|\Delta-2|\right).
\end{equation}
In the case of twist 2 operators, $\Delta=J+2$, one has $\xi=J$ and
the relation~\eqref{14} reproduces the Regge spectrum of original SW model~\cite{son2},
$m_{n,J}^2=4a(n+J)$.

\section{Discussions and conclusion}

The case $\Delta=2$ in~\eqref{14} formally leads to a massless ground state. This observation
gave rise to suggestions on interpretation of the corresponding scalar meson as the physical pion
(see, e.g.,~\cite{br3,afonin2012,br4,gutsche};
strictly speaking, these suggestions were proposed in the case $a<0$). It must be emphasized that
the mass relation~\eqref{14} is not valid for scalar fields corresponding to
scaling dimension $\Delta=2$ as the potential~\eqref{12}
is not bounded from below in this case. As is known from Quantum Mechanics~\cite{landau}, the critical value
$\xi=0$ in~\eqref{12} corresponds to the onset of "falling towards the center" when
the descrete spectrum (even with negative eigenvalues) disappears.

Within the light-front holographic QCD~\cite{br2,br3}, the quantity $\xi$ in~\eqref{6}
represents the light-front internal orbital angular momentum $L$.
The scaling dimension $\Delta$ and $L$ are simply related due to~\eqref{3},
$\Delta=2+L$. The spectrum~\eqref{13} leads then to a remarkable relation,
\begin{equation}
\label{15}
m_{n,L}^2\sim n+L.
\end{equation}
If $L$ is identified with the real orbital angular momentum
of quark-antiquark pair in mesons, the relation~\eqref{15} predicts
a hydrogen-like kind of degeneracy in meson spectra~\cite{hydrogen}. The spectral law~\eqref{15}
was observed in the light non-strange mesons by several authors~\cite{deg1,deg2}.
We get thus a qualitative holographic description of this striking phenomenon for $L>0$.

The SW meson spectrum which is believed to follow from the light-front holographic QCD is different:
$m_{n,L,J}^2\sim n+(L+J)/2$~\cite{br3,gutsche} (the relation~\eqref{15}, however,
was obtained in the original paper~\cite{br2}). The difference emerges
by virtue of absence of the "mass renormalization"~\eqref{3b}. Without this 
infrared renormalization of 5D mass the discrete spectrum is
\begin{equation}
\label{18}
m_{n,J}^2=2|a|\left[2n+1+\xi-\frac{a}{|a|}(J-1)\right].
\end{equation}
In this case, the acceptable dependence of mass on spin requires the opposite sign
of dilaton background, $a<0$~\cite{br3,afonin2010}. With negative $a$ and identifying $\xi$ with $L$
we reproduce the spectrum obtained in Refs.~\cite{br3,gutsche}, $m_{n,L,J}^2=4|a|\left[n+(L+J)/2\right]$.
The splitting of states with different $L$ looks like an attractive feature of this spectrum.
For instance, the vector and axial spectra behave as $m_{n,0,1}^2\sim n+1/2$
and $m_{n,1,1}^2\sim n+1$, correspondingly
(note that these relations were known in the planar QCD sum rules~\cite{afonin2003}).
Also the massless ground state having $J=L=0$ was interpreted as the pion. As was remarked above,
we disagree with the given statement since the discrete spectrum is absent in this case.

For description of the $L=0$ vector states in~\eqref{15}
we need some additional assumptions.
On a phenomenological level, one could assume for instance that the $SO(2)$ light-front
angular momentum $L_{\text{l-f}}$ and the $SO(3)$ physical angular momentum $L_{\text{ph}}$
are related via some shift, $L_{\text{l-f}}=L_{\text{ph}}+b$.
In order to estimate the shift $b$ let us identify $\xi=L_{\text{l-f}}$ at which
a massless eigenvalue appears in Eq.~\eqref{11} with a physical $S$-wave state, $L_{\text{ph}}=0$.
As follows from the potential~\eqref{12}, the value $\xi=1/2$ corresponds to the spectrum
of one-dimensional oscillator, $m_n^2=(2n+1)|a|-a$. As long as we consider $a>0$,
the ground state indeed has zero energy. We get thus $b=1/2$ and
the physical spectrum $m_{n,L}^2\sim n+L_{\text{ph}}+1/2$. The additional
contribution $1/2$ could be also interpreted as arising from transition to three
dimensions (real space) from two dimensions (light front) as it follows from the spectrum of linear
oscillator, $E_n=\hbar\omega(n+d/2)$, where $d$ is the space dimension.
It is interesting to note
that the phenomenological value of intercept $b$ seems to be indeed
close to $1/2$ in the light mesons~\cite{deg1}.

\section*{Acknowledgments}

The work was partially supported by the RFBR grant 16-02-00348-a.

\end{document}